\documentclass[
  ,final 
  ] {aipproc}

\layoutstyle{6x9}
\newcommand{\teff}{T_{\rm eff}}
\newcommand{\bq}{\begin{eqnarray}}
\newcommand{\eq}{\end{eqnarray}} 

\begin{document}

\title{$\teff$ and $\log{g}$ dependence of velocity fields in M-stars}


\author{S.~Wende}{
  address={Institut f\"ur Astrophysik, Georg-August-Universit\"at G\"ottingen,
  Friedrich-Hund Platz 1, D-37077 G\"ottingen, Germany} }

\author{A.~Reiners}{
  address={Institut f\"ur Astrophysik, Georg-August-Universit\"at G\"ottingen,
  Friedrich-Hund Platz 1, D-37077 G\"ottingen, Germany} }

\author{H.-G.~Ludwig}{
  address={GEPI, CIFIST, Observatoire de Paris-Meudon, 5 place Jules Janssen,
  92195 Meudon Cedex, France} }

\begin{abstract}
  We present an investigation of velocity fields in early to late M-type
  hydrodynamic stellar atmosphere models. These velocities will be expressed
  in classical terms of micro- and macro-turbulent velocities for usage in 1D
  spectral synthesis. The M-star model parameters range between $\log{g}$ of
  $3.0$ -- $5.0$ and $\teff$ of $2500$~K -- $4000$~K. We characterize the
  $\teff$- and $\log{g}$-dependence of the hydrodynamical velocity fields in
  these models with a \emph{binning} method, and for the determination of
  micro-turbulent velocities, the \emph{Curve of Growth} method is used. The
  macro-turbulent velocities are obtained by convolutions with Gaussian
  profiles.  Velocity fields in M-stars strongly depend on $\log{g}$ and
  $\teff$. Their velocity amplitudes increase with decreasing $\log{g}$ and
  increasing $\teff$.  The 3D hydrodynamical and 1D macro-turbulent velocities
  range from $\sim 100$~m/s for cool high gravity models to $\sim
  800$~m/s--$1000$~m/s for hot models or models with low $\log{g}$. The
  micro-turbulent velocities range in the order of $\sim 100$~m/s for cool
  models, to $\sim 600$~m/s for hot or low $\log{g}$ models. Our M-star
  structure models are calculated with the 3D radiative-hydrodynamics (RHD)
  code \texttt{CO$^5$BOLD}. The spectral synthesis on these models is
  performed with the line synthesis code \texttt{LINFOR3D}.
\end{abstract}

\maketitle


\section{Introduction}
The measurement of line broadening in cool stars is in general a difficult
task. For example, the investigation of the rotation-activity connection among
field M-dwarfs requires the measurement of rotational line broadening with an
accuracy of $1$~km/s \citep{2007A&A...467..259R}. The spectral lines have to
be very narrow and well isolated to detect slow rotation. But in cool stars
most of the individual atomic lines become very weak at these low temperatures
and are dominated by pressure broadening. Also the measurement of the magnetic
field strength is dependent on the line width, and detection of Zeeman
splitting becomes more difficult at low temperatures due to the aforementioned
reasons. Since it is possible to use the narrow and well isolated FeH molecule
lines in cool stars to determine radial velocities or magnetic field strength
\cite{2007ApJ...656.1121R}, it would be very helpful to characterize the
impact of macroscopic velocity fields -- primarily driven by convection -- on
the line shapes from hydrodynamical model atmospheres.  We calculate
3D-\texttt{CO$^5$BOLD} stellar atmosphere models \citep{2002A&A...395...99L}
which serve as input to the line formation program \texttt{LINFOR3D}
\citep[based on][]{Bascheck1966}. Velocity fields and thermal inhomogeneity
are naturally represented in the hydrodynamical models. The influence on the
modeled spectral lines can then be investigated and translated into effective
micro- and macro-turbulent velocities used in classical analyses based on 1D
hydrostatic model atmospheres. The comparison with 1D-models provides some
insight when it is necessary to apply 3D-models in the spectral analysis of
cool stars.

In the first part of the paper we describe our hydrodynamical model atmosphere
code and give an overview of the models, in the second part we investigate the
velocity fields in the models and their dependence on $\log{g}$ and $\teff$.
We will express the influence of velocity fields on line shapes in terms of
micro- and macro-turbulent velocities.

\section{Methods and Models}
\texttt{CO$^5$BOLD} is the abbreviation for ``COnservative COde for the
COmputation of COmpressible COnvection in a BOx of L Dimensions with L=2,3''
\citep{2008asd..soft...36F}.  It can be used to model solar and stellar
surface convection. In solar-like stars, a \texttt{CO$^5$BOLD} model
represents the 3D flow geometry and its temporal evolution in a small
(relative to the star's radius) Cartesian domain at the stellar surface (``box
in a star'' set-up). The spatial size of the domain is chosen to be sufficient
to include the dominant convective scales, i.e. the computational box is large
enough to include a number of granular cells at any instant in time. A
\texttt{CO$^5$BOLD} model provides a statistical realization of the convective
flow. In this investigation we usually average over five flow fields taken at
different instances in time (``snapshots'') to improve the statistics.

\texttt{CO$^5$BOLD} solves the coupled non-linear time-dependent equations of compressible
hydrodynamics coupled to the radiative transfer equation in an external
gravitational field in 3 spatial dimensions. As set of
independent quantities are chosen the mass density $\rho$, the three
spatial velocities $v_x$, $v_y$, and $v_z$, and the internal energy
$\epsilon_{i}$. With these quantities, the 3D hydrodynamics equations,
including source terms due to gravity, are the mass conservation equation
\bq
\frac{\partial \rho}{\partial t}+\frac{\partial \rho v_x}{\partial \rho x}
                                +\frac{\partial \rho v_y}{\partial \rho y}
                                +\frac{\partial \rho v_z}{\partial \rho z}=0 ,
\eq
the momentum equation
\bq
\frac{\partial}{\partial t} 
\left( \begin{array}{c} \rho v_x \\ \rho v_y \\ \rho v_z \end{array} \!
\right) +
\frac{\partial}{\partial x}
\left( \begin{array}{l} \rho v_x v_x +P\\ \rho v_y v_x\\ \rho v_z v_x\end{array} \!
\right) +
\frac{\partial}{\partial y}
\left( \begin{array}{l} \rho v_x v_y \\ \rho v_y v_y + P\\ \rho v_z v_y\end{array} \!
\right) +
\frac{\partial}{\partial z}
\left( \begin{array}{l} \rho v_x v_z \\ \rho v_y v_z \\ \rho v_z v_z +P\end{array} \!
\right)=\left( \begin{array}{c} \rho g_x \\ \rho g_y \\ \rho g_z \end{array}
 \right),
\eq
and the energy equation which includes the radiative heating term $Q_{rad}$
\bq
\frac{\partial \rho \epsilon_{ik}}{\partial t}+\frac{\partial (\rho \epsilon_{ik}+P)v_x}{\partial x}
+\frac{\partial (\rho \epsilon_{ik}+P)v_y}{\partial y} +\frac{\partial (\rho
\epsilon_{ik}+P)v_z}{\partial z} =\rho(g_x v_x+g_y v_y+g_z v_z)+Q_{rad}.
\eq
$\epsilon_{ik}$ denotes the sum of internal and kinetic energy.
The gas pressure P is related to density $\rho$ and internal energy
$\epsilon_i$ via a (tabulated) equation of state $P=P(\rho,\epsilon_i)$.
For the local models used here the gravity field is given by the constant vector
$\vec{g}=\left(\begin{array}{r} 0\\0\\-g \end{array}\right)$.
\texttt{CO$^5$BOLD} uses the convention that the vertical axis points upwards.
The radiative heating term $Q_{rad}$ is obtained from the solution of the
non-local frequency-dependent radiative transfer equation. The frequency
dependence of the radiation field is captured by considering a small number
of representative  wavelength bands 
\citep[``opacity binning'', see ][]{2002A&A...395...99L,2006A&A...459..599L}.
The resulting 3D radiative-hydrodynamic (RHD) models treat convection from
basic physical principles and avoid approximations like mixing-length theory.
In the following, the three dimensional data cubes of the \texttt{CO$^5$BOLD}-models will be called
``3D-models'', and the spectral lines computed from three dimensional
atmosphere models ``3D-lines''. 

We will characterize the velocity fields in
the 3D RHD models and analyze their influence of FeH lines. For the 1D
spectral synthesis the 3D velocity fields will be expressed in terms of micro-
and macro-turbulent velocity. For this investigation, we choose a set of
\texttt{CO$^5$BOLD}-models with $\teff=2500$~K -- $4000$~K and
$\log{g}=3.0$ -- $5.0$ [cgs]. Table \ref{tab1} gives the model parameters. 
\begin{table}[!h]
\centering
\begin{tabular}{lrrrrr}
\hline\hline
 Model code & Dim. & Size(x,y,z) [km] & Opacities & $\teff$ [K] & $\log{g}$
 [cgs]\\
\hline
d3t33g30mm00w1 & 3 & 85000 x 85000 x 58350 & PHOENIX & 3240 & 3.0 \\
d3t33g35mm00w1 & 3 & 28000 x 28000 x 11500 & PHOENIX & 3270 & 3.5 \\
d3t33g40mm00w1 & 3 & 7750 x 7750 x 1850 & PHOENIX & 3315 & 4.0 \\
d3t33g50mm00w1 & 3 & 600 x 600 x 260 & PHOENIX & 3275 & 5.0 \\
d3t40g45mm00n01 & 3 & 4700 x 4700 x 1150 & PHOENIX & 4000 & 4.5 \\
d3t38g49mm00w1 & 3 & 1900 x 1900 x 420 & PHOENIX & 3820 & 4.9 \\
d3t35g50mm00w1 & 3 & 1070 x 1070 x 290 & PHOENIX & 3380 & 5.0 \\
d3t28g50mm00w1 & 3 & 370 x 370 x 270 & PHOENIX & 2800 & 5.0 \\ 
d3t25g50mm00w1 & 3 & 240 x 240 x 170 & PHOENIX & 2575 & 5.0 \\
\hline
\end{tabular}
\caption{Overview of different model quantities. We simulated main sequence
stars and varied $\log{g}$ slightly for models with changing $\teff$. The
models with almost same $\teff$ were started at the same $\teff$ of $3300$~K,
but they converge at slightly lower or higher $\teff$.}
\label{tab1}
\end{table}
The opacities used in all models are obtained from the \texttt{PHOENIX}
stellar atmosphere package \citep{1999JCoAM.109...41H} assuming an abundance
mixture according \citet{agsSolAbun}. The opacity tables were computed after
\citet{jwfOpac05} and \citet{bdCO5BOLD}.

In order to compare 1D model atmospheres with the hydrodynamical 3D-models, we
average the 3D-models over surfaces of equal optical depth. We obtain
so-called $<$3D$>$-models which have the same mean thermal profile as in the
3D-models but without cold and hot regions which stem from the convective
granulation pattern. In the $<$3D$>$-models the hydrodynamic velocity field is
not considered. The line-broadening is treated in the classical way by adding
isotropic Gaussian micro- and macro-turbulence. We will call the spectral
lines of these models $<$3D$>$-lines.

\section{Hydrodynamical velocity fields}
\begin{figure}[!h]
\centering
 \includegraphics[width=7cm,bb=19 29 580
 410]{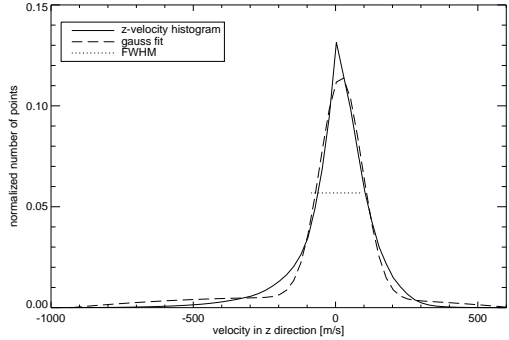} \caption{Histogram of the
 velocity distribution in vertical direction. The normalized number of points
 is plotted against the vertical velocity in m/s (solid line). 
 The Gaussian (dashed line) fits the velocity distribution and
 determined an FWHM value (dashed-dotted line), which is related to $\sigma$
 with $FWHM=2\sqrt{2\ln{2}}\cdot\sigma$. The
 underlying model is located at $\teff=2800$~K and $\log{g}=5$~[cgs]. }
\label{FWHMhisto}
\end{figure}

\begin{figure}[!h]
\centering
 \includegraphics[width=7cm]{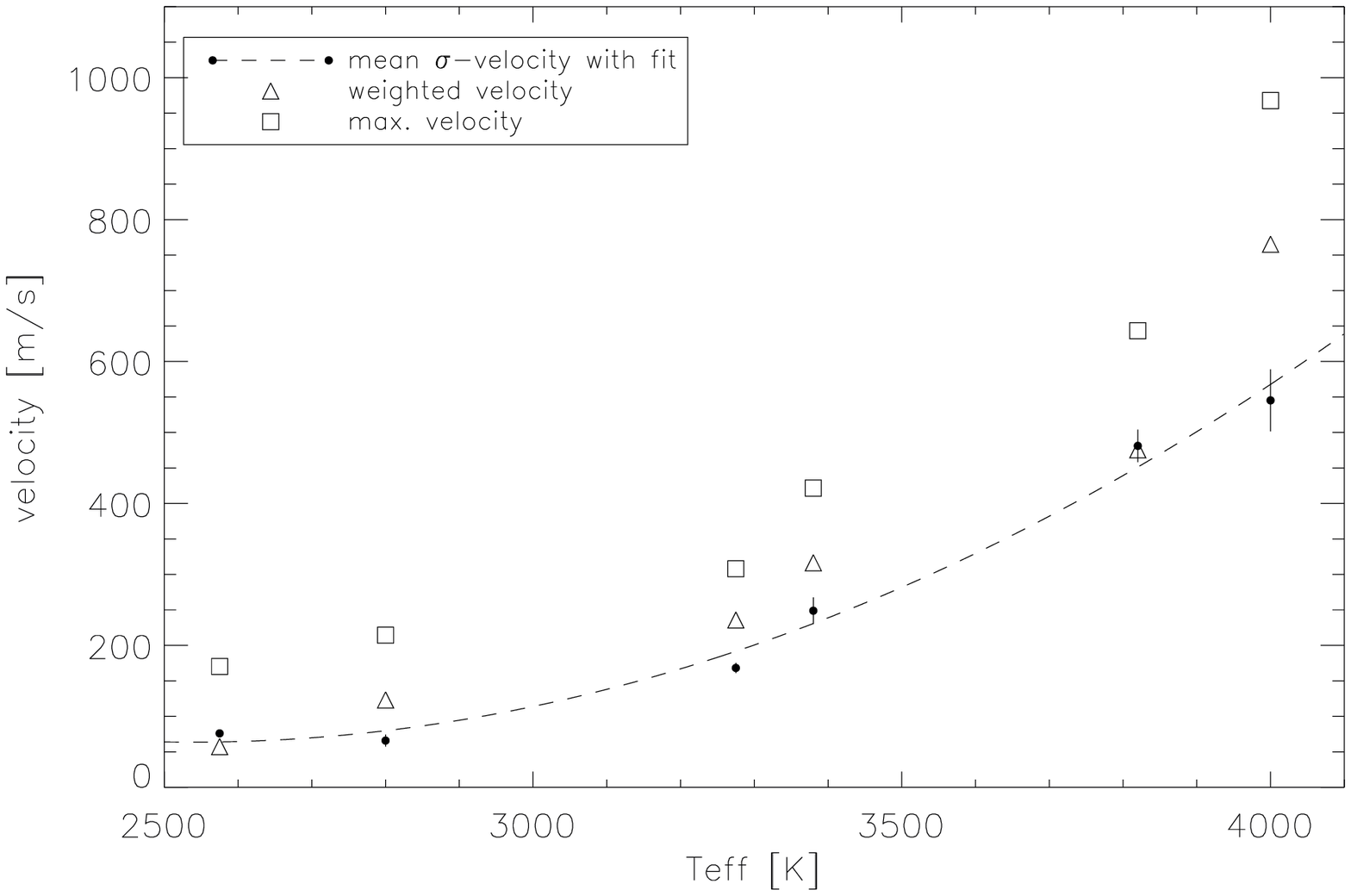}
 \includegraphics[width=7cm]{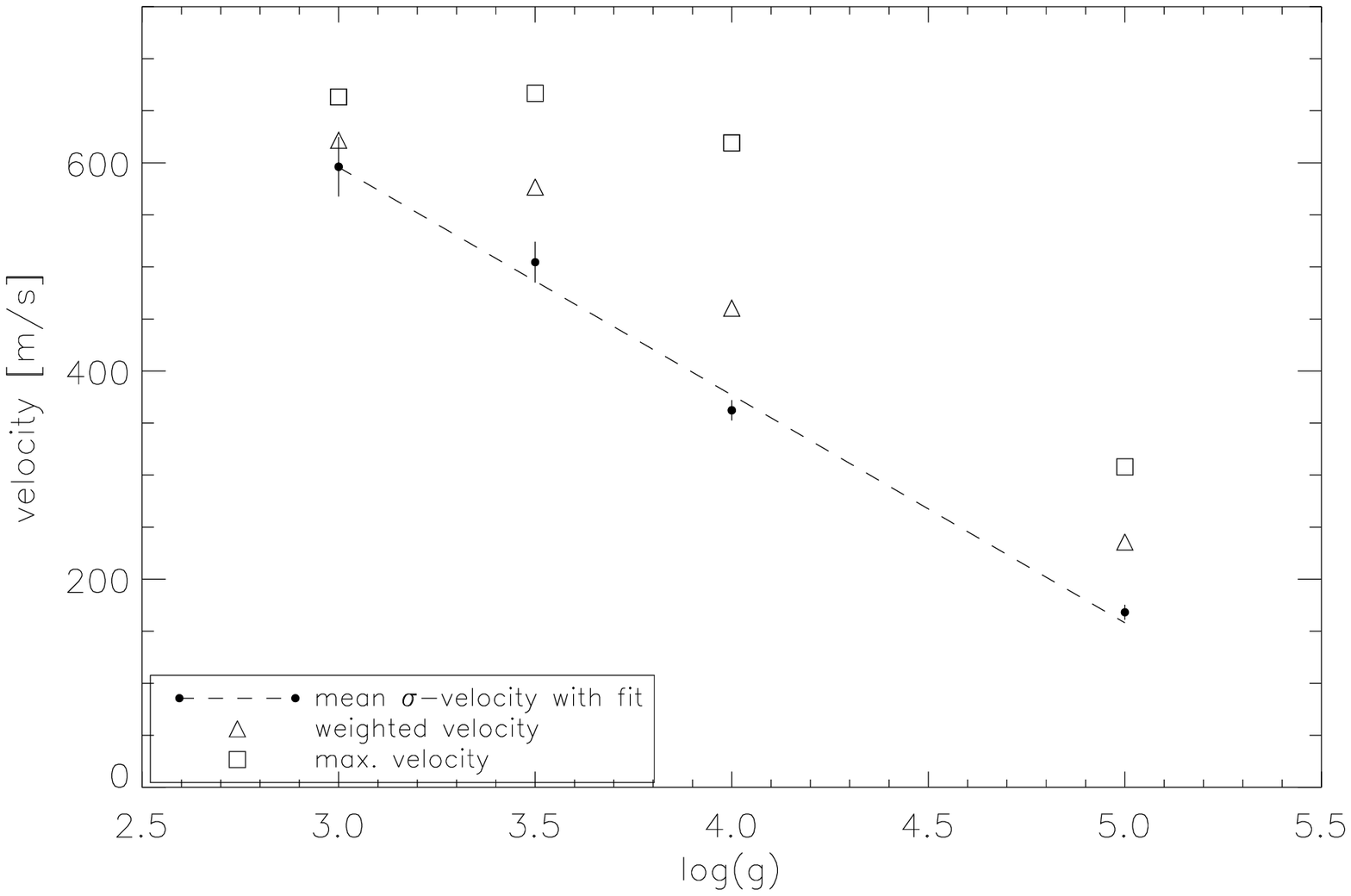}
\caption{Plotted are
 velocity dispersions with error bars, maximal velocities in the convection
 zone, and weighted velocity dispersions (see text) for models with different
 $\teff$ (left) and different $\log{g}$ (right). The dashed line on the left
 plot shows a second order polynomial fit for the velocity dispersion and on
 the right plot a linear function is sufficient to fit the $\log{g}$ data.}
\label{absvelos}
\end{figure}

\begin{figure}[!h]
\centering
 \includegraphics[width=14cm,bb=25 30 530
 225]{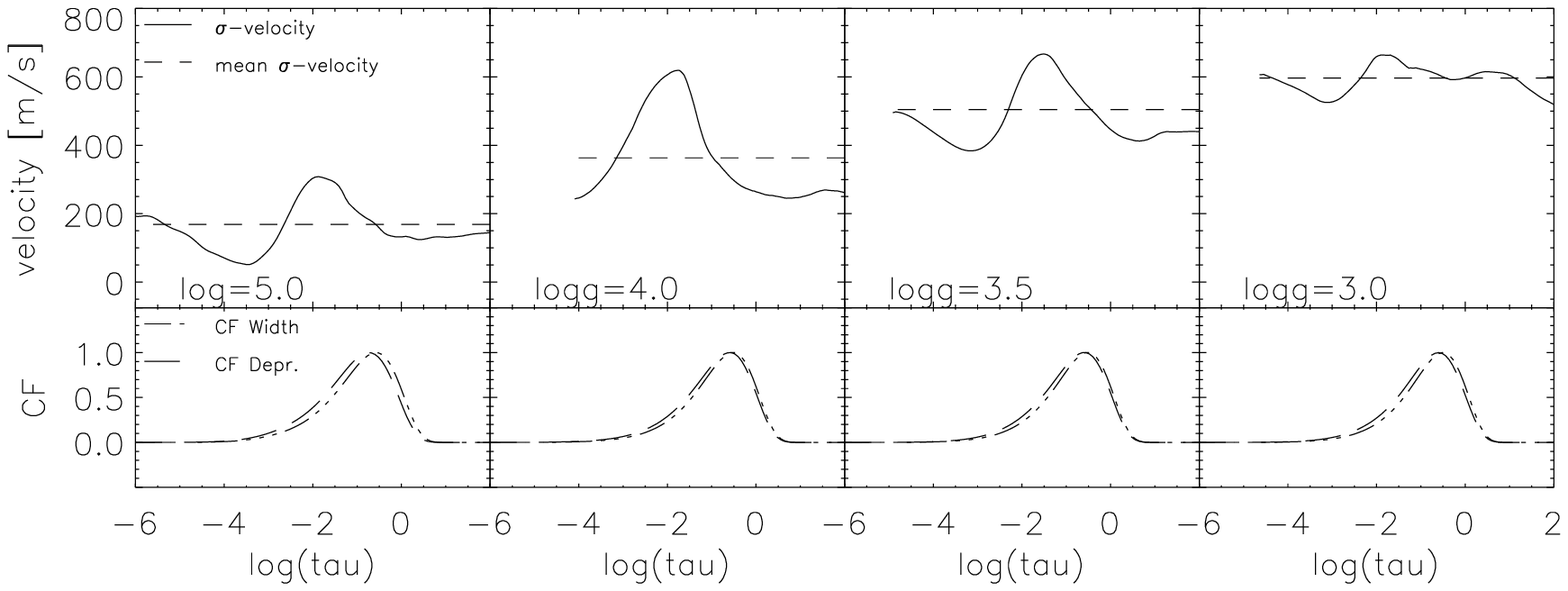} 
\caption{Upper panel: $\sigma$
 velocities.In each upper panel the $\sigma$ -velocities (solid) and the mean
 $\sigma$-velocity (dotted) are plotted against the optical depth in logarithmic
 scale. Bottom panel: In each bottom panel are the contribution functions (CF)
 of a FeH-line at a wavelength of $9954.0$~\AA\. The width (dashed triple
 dotted) and depression (dashed) of the line as a function of optical depth on
 a logarithmic scale.}
\label{sigma}
\end{figure}

Spectral lines are broadened by velocity fields where the wavelength of
absorption or emission of a particle is shifted due to its motion in the gas.
Here we are mostly concerned with the macroscopic, hydrodynamic motions but
have in mind that the thermal motions are also constituting an important
contribution.  If we envision each voxel in the RHD model cube to form its own
spectral line, the whole line consists of a (weighted) sum of single lines.
The velocity distribution might be represented by a histogram of the
velocities of the voxels. We try to describe the velocity fields in that sense
instead of using the rms-velocities. Since a velocity vector is assigned to
each voxel, we apply a binning method and count all vertical velocities to
plot them in a histogram with a bin size of $25$~m/s. We took the standard
deviation $\sigma$ of a fitted Gaussian normal distribution as a measure for
the velocity dispersion in the models (see Fig.\ref{FWHMhisto}). At this
stage of the investigation time we only concentrate on the vertical direction
which is appropriate if we assume that the major part of the line broadening
stems from the vertical motions. The velocities of the horizontal directions
are of roughly the same order.
As a measure for the statistical variability in the 3D RHD model, we average over
five different temporal windows and also compute the standard deviation. The
determined velocities are plotted against $\teff $ and $\log{g}$ to
investigate their dependence (Fig.~\ref{absvelos}). The velocities increase
strongly with decreasing $\log{g}$ and with increasing $\teff$. They are as
low as a few hundred m/s in the coolest high-gravity models. We can
describe the dependence of the velocity dispersions on $\teff$ with a second
order polynomial, and in $\log{g}$ with a linear function (The fits for
the velocity dispersion are overplotted in Fig.\ref{absvelos}).\\ In order to
investigate in which region of the model atmospheres the velocity fields are
generated and what the physical meaning of the $\sigma$ is, we bin the 3D model
velocities at constant geometrical height and fit a Gaussian to determine the
velocity dispersion
for each layer (Fig.~\ref{sigma}).  
Typically, we find maxima of the velocity dispersion around $\log{\tau}~=~-1$.
The maximum velocities in the
convection zone of each model ($\teff$ and $\log{g}$) are plotted in
Fig.~\ref{absvelos} too.  
\begin{figure}[!h]
\centering
 \includegraphics[width=14cm,bb=25 30 530
 225]{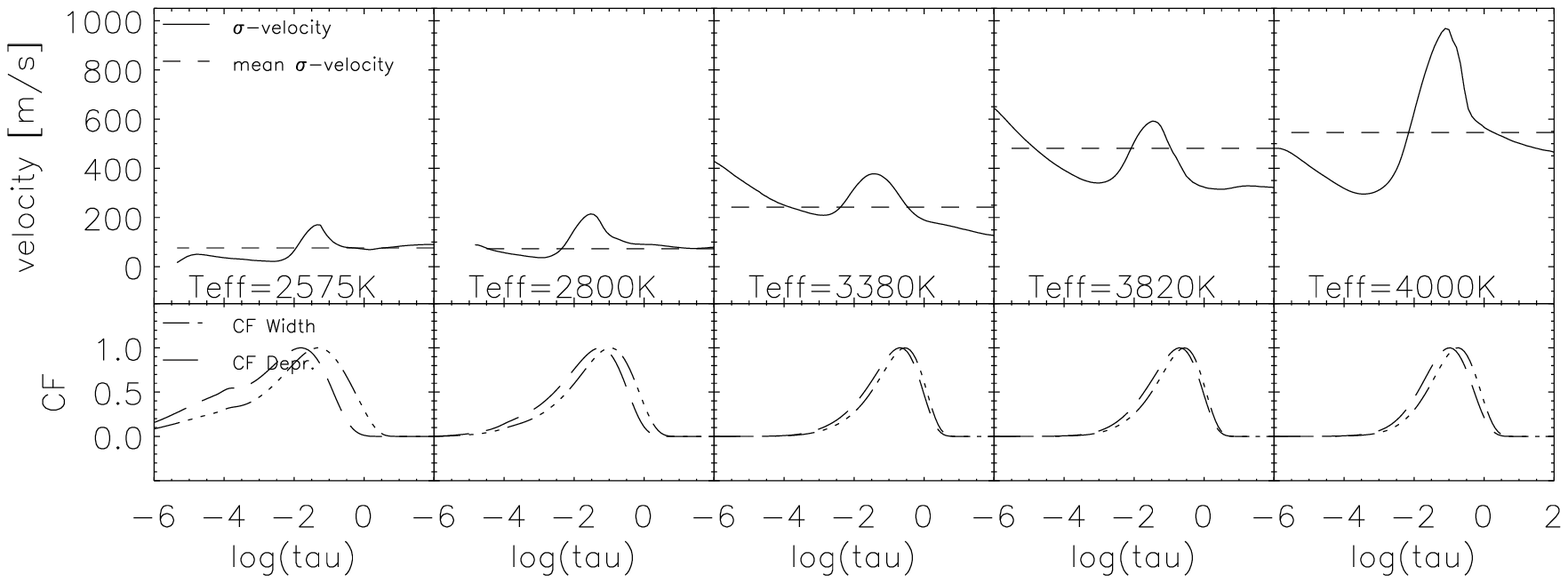}
\end{figure}
In order to measure the velocities in the region where the lines originate, we
compute the mean of the velocities weighted with the contribution function of
the line depression \citep{1986A&A...163..135M}
$\sigma_{weighted}=\frac{\sum_{\tau=2}^{-6}\sigma_{\tau}\cdot
CF_{\tau}}{\sum_{\tau=2}^{-6}CF_{\tau}}$ showed in the bottom panel of
Fig.~\ref{sigma}.  These velocities are also plotted in
Fig.~\ref{absvelos}. In most cases they lie between the velocity dispersion
and the maximum velocity in the convection zone. 
Some models show a strong increase of the vertical velocity in higher layers
(Fig.~\ref{sigma}). This is related to the presence of waves which are excited
by the stochastic fluid motions
\citep[][and references therein]{2002A&A...395...99L}.
However, it will not affect the spectral lines, because they are generated in
the region between an optical depth of $\log{\tau}~=~1.0$ and $\log{\tau}~=~-4.0$.

\begin{figure}[!h]
\centering
 \includegraphics[width=8cm,bb=20 20 606 386]{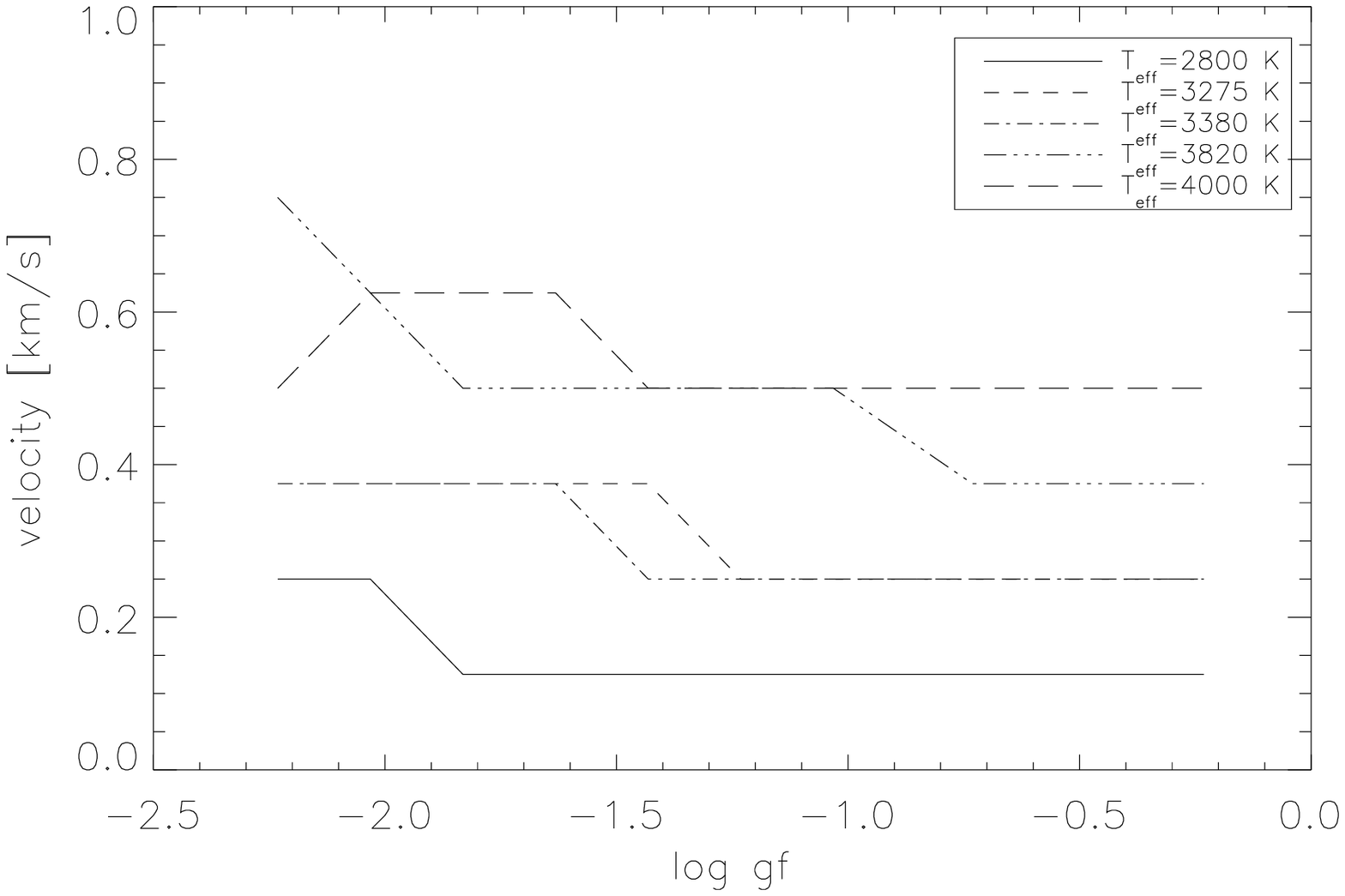}
 \includegraphics[width=8cm,bb=20 20 606 386]{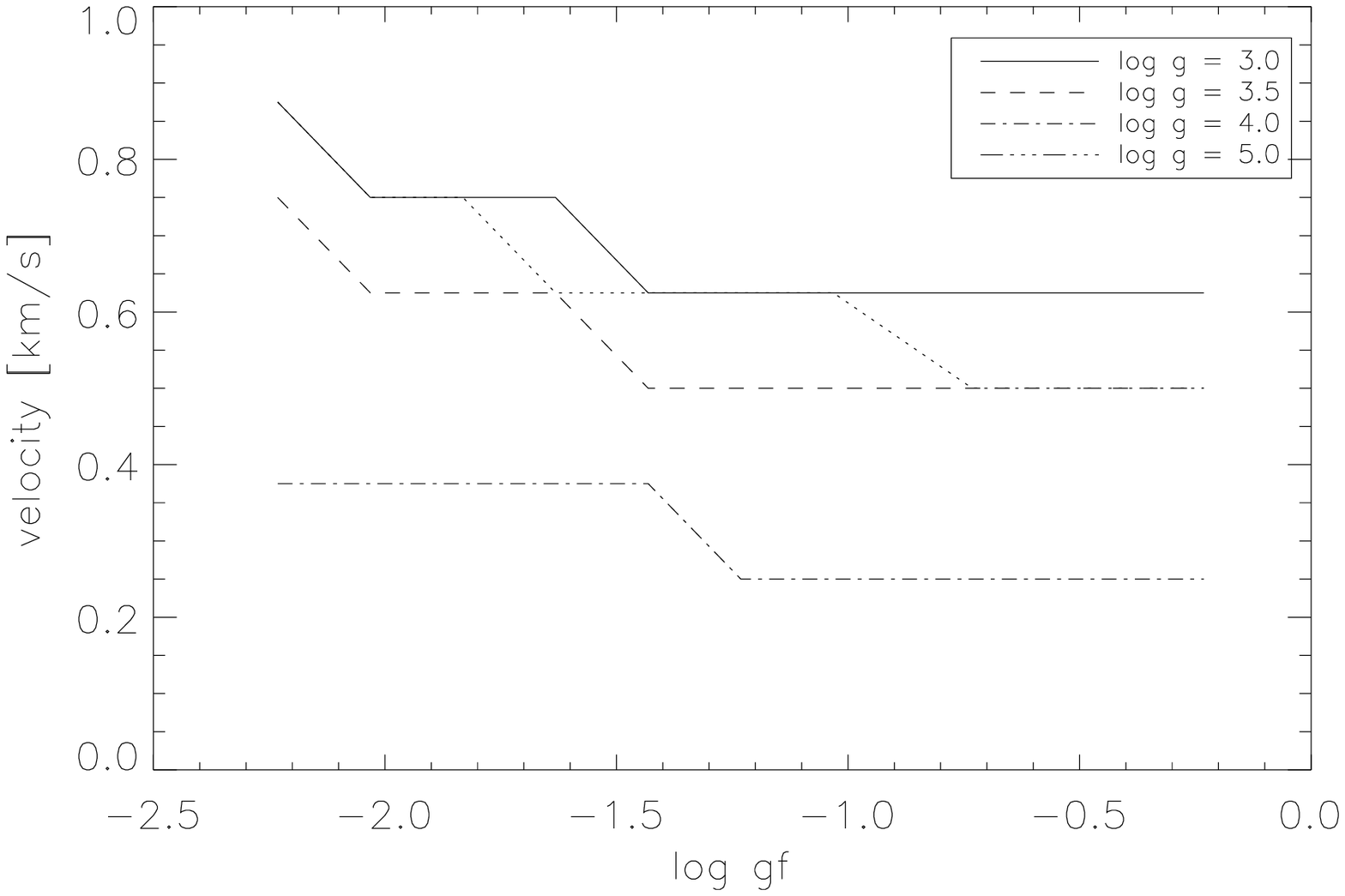}
 \caption{Micro-turbulent velocities as a function of $\log{gf}$ for different $\teff$
 (left) and different $\log{g}$ (right).}
\label{loggfvelos}
\end{figure}
\begin{figure}[!h]
\centering
 \includegraphics[width=8cm]{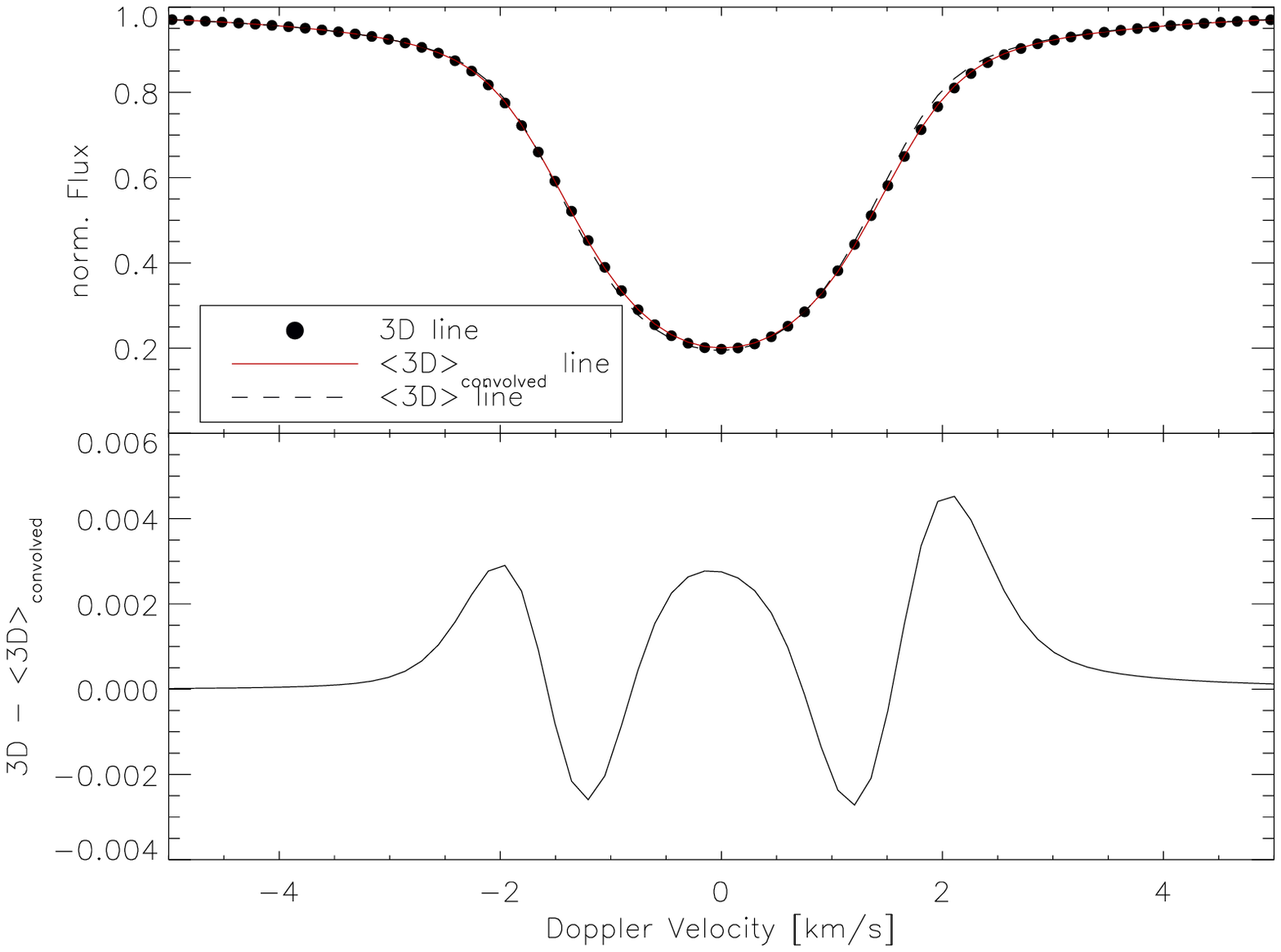}
 \includegraphics[width=8cm]{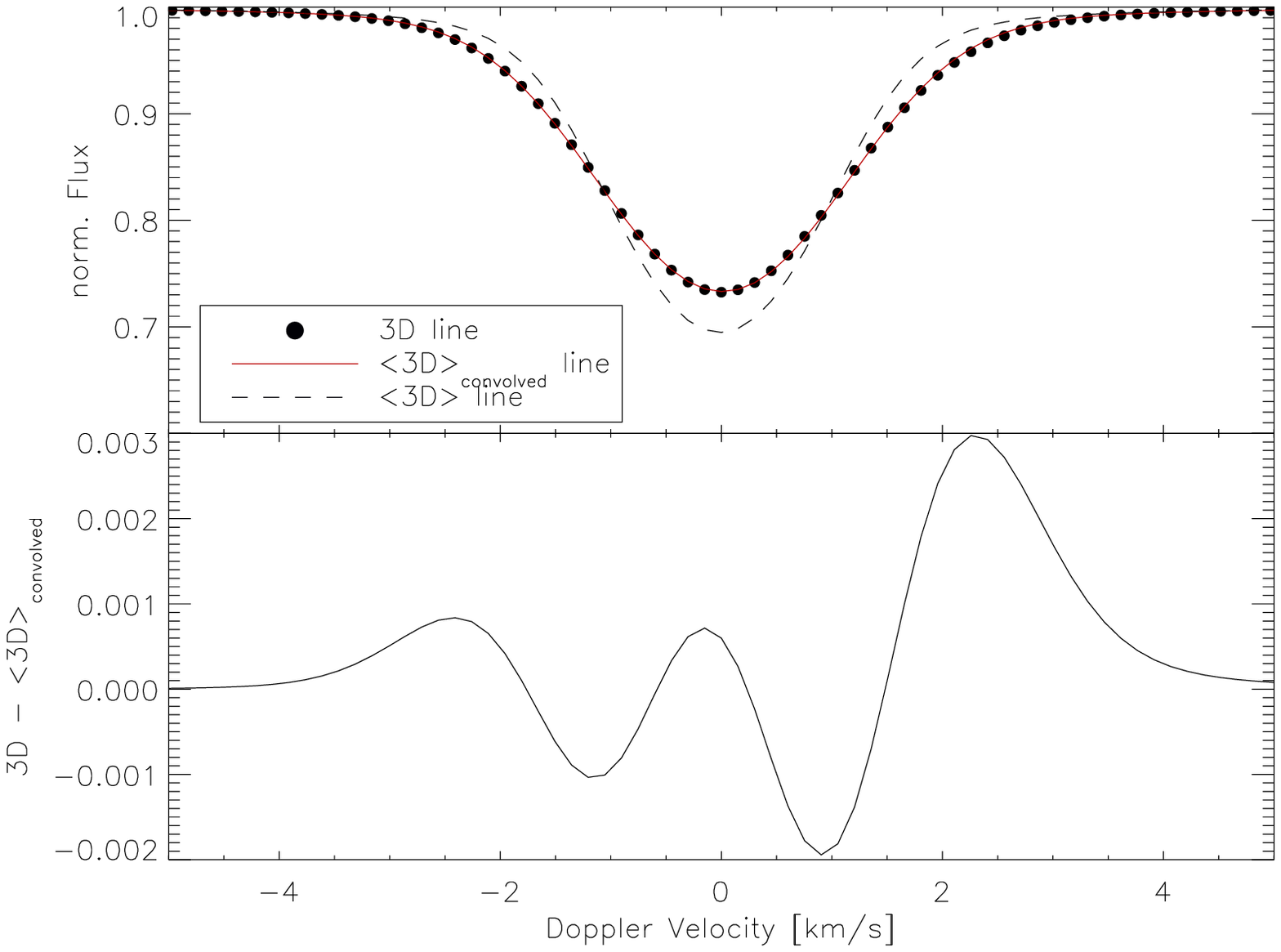}
 \caption{FeH lines for models with $\teff~=~2800$~K, $\log{g}~=~5.0$ [cgs]
 (left) and $\teff~=~3820$~K, $\log{g}~=~4.9$ [cgs] (right).  The upper panels
 show the 3D-line (dots) and the $<$3D$>$$_{convolved}$-line (solid line)
 which was broadened by a Gaussian profile. For comparison we plotted a
 $<$3D$>$-line which was not broadened by any velocities (dashed line). In the
 lower panels are the 3D-$<$3D$>$$_{convolved}$ residuals plotted. One can see the asymmetry which
 stems from the line shifts due to convective motions. }
\label{gausconv}
\end{figure}
   
\begin{figure}[!h]
\centering
 \includegraphics[width=8cm,bb=20 20 606 386]{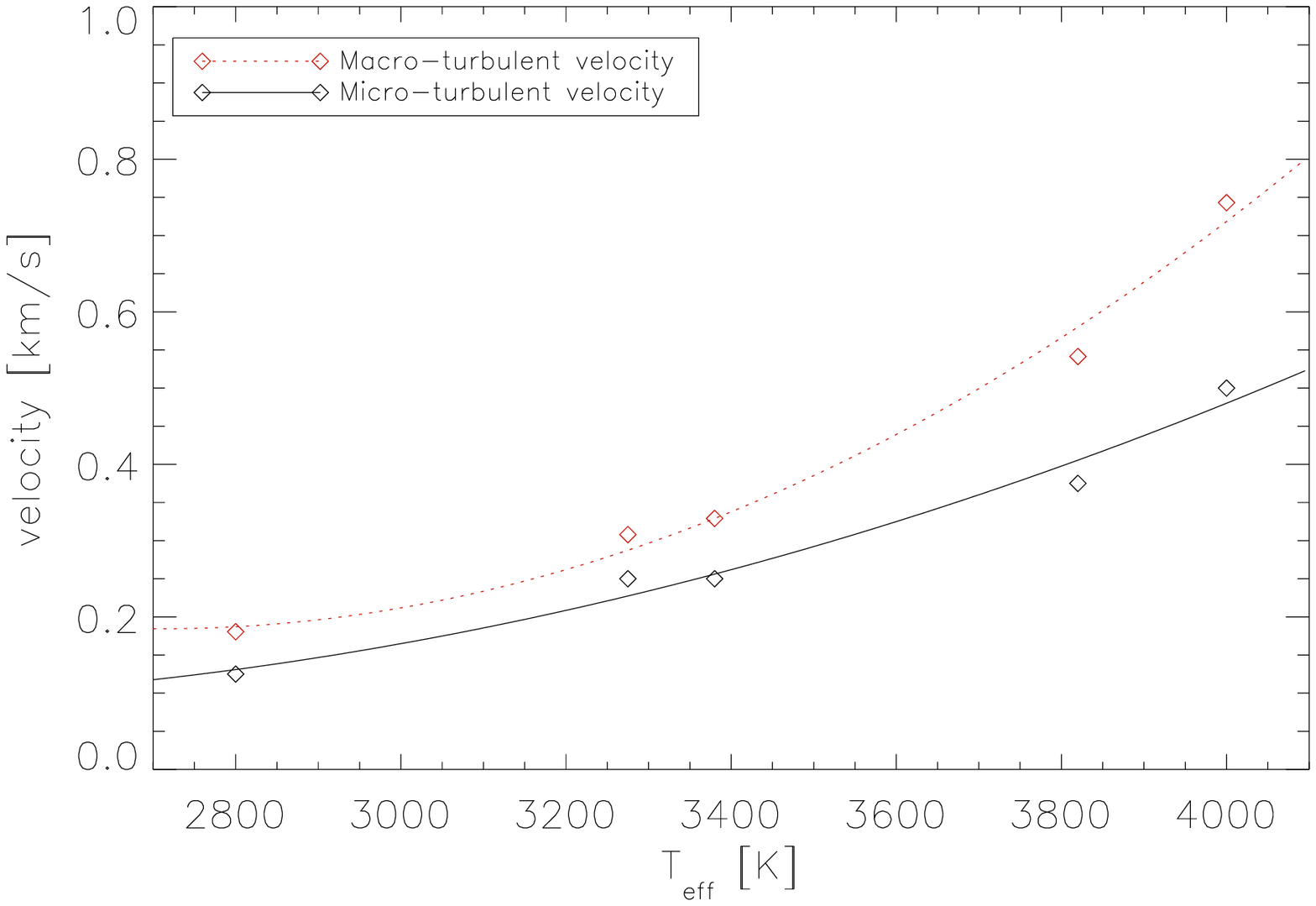}
 \includegraphics[width=8cm,bb=20 20 606 386]{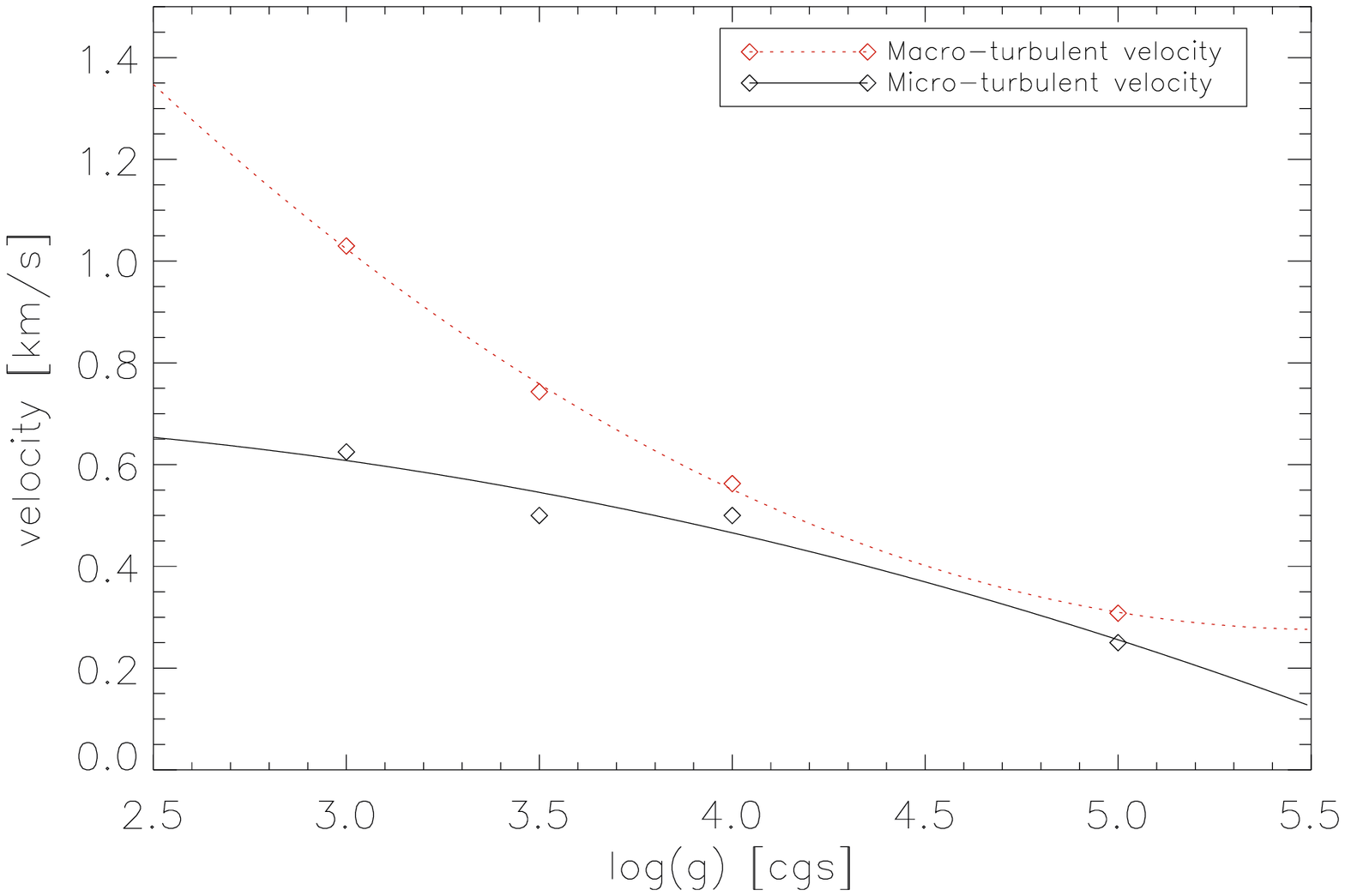}
 \caption{Micro- and macro-turbulent velocities as a function of $\teff$
 (left) and $\log{g}$ (right). The data points are fitted by second order
 polynomials.}
\label{mmvelos}
\end{figure}
\clearpage
\section{Micro- and macro-turbulent velocities}
In the following we express the 3D hydrodynamical velocities in terms of
classical micro- and macro-turbulent velocities \citep[see
e.g.][]{1977ApJ...218..530G,2008oasp.book.....G}. We try to see how accurately
the 1D broadening resembles the 3D broadening in the regime of cool stars
which, as we saw, exhibit velocity fields of relatively small amplitudes. A 1D
treatment would save a lot of CPU-time, and little differences between 1D and
3D broadening could favor the usage of fast 1D atmosphere codes to simulate
M-stars for comparisons with observations, e.g. to determine rotational- or
Zeeman-broadening.\\
{\bf\emph{Micro-turbulent velocities:}} In order to determine the
micro-turbulent velocities, we use the \emph{curve of growth} (CoG) method
\citep{2008oasp.book.....G}. We increase the line strength of an FeH and an
FeI line by (artificially) increasing the $\log{gf}$ value of the line which
results in an increasingly saturated lines. In saturated lines, the effective
absorbing cross-section is enhanced due to the influence of the
micro-turbulent velocities. From 3D spectral synthesis we obtain a 3D CoG,
which can we compare with the $<$3D$>$ CoG for different micro-turbulent
velocities which are included in the 1D spectral synthesis. We use a grid of
micro-turbulent velocities that range from $0$~km/s to $1$~km/s in
$0.125$~km/s steps and obtain $<$3D$>$ CoGs for nine different micro-turbulent
velocities.  We compare the $<$3D$>$ CoGs with the 3D CoG and select the
velocity of the $<$3D$>$ CoG that fits the 3D CoG best. The determined
velocities are plotted in Fig.\ref{mmvelos}. Since strong (saturated) lines
tend to be formed in the upper atmosphere, and weak (unsaturated) lines in
deeper layers of the atmosphere, $\log{gf}$ and the height of formation are
related. If we fit a micro-turbulent velocity to each $\log{gf}$ point of the
3D CoG, we obtain a height-dependent velocity structure (see Fig.\ref{loggfvelos}).\\
{\bf \emph{Macro-turbulent velocities:}} For modeling macro-turbulent
broadening we used the radial-tangential profile from
\citet{1975ApJ...202..148G}, as well as a Gaussian profile in the 1D spectral
synthesis. We compared $<$3D$>$ absorption lines which were broadened with
both profiles, and the results differ only very little. Hence it is
appropriate to use the Gaussian profile here. It is
remarkable that the broadening due to small 3D RHD velocity fields in M-stars
can be approximated with a simple Gaussian velocity distribution. Two examples
of Gaussian broadened $<3D>$ FeH lines are shown in Fig.~\ref{gausconv}. To
determine the macro-turbulent velocities of the 3D RHD models, we first
compute the $<3D>$ lines with a given micro-turbulent velocity and after this
we broaden them with a Gaussian with a given macro-turbulent velocity until
they match the 3D profiles. The results are plotted in Fig.\ref{mmvelos}.
The broadened $<3D>$ FeH lines fit the 3D FeH lines very well. The difference
of the $<3D>$ and 3D centroid ($C=\frac{\sum F\cdot v}{\sum F}$) range in the
order of a few m/s for small velocity fields to $30-40$~m/s for strong
velocity fields in hot M star models, or models with low $\log{g}$. The differences
stem from the asymmetry of the 3D line profiles, which are shifted due to convective
motions. However, the errors in the normalized flux profiles are lower than $1\%$ (see Fig.~\ref{gausconv}).
We fit the micro- and macro-turbulent velocities as a function of $\teff$ and
$\log{g}$, each pair can be fitted with a polynomial second order. The fits
are overplotted in Fig.\ref{mmvelos}.  Both micro- and macro-turbulent
velocities show similar dependence on $\teff$ and $\log{g}$ as the RHD
velocity fields considered before (i.e. they increase with increasing $\teff$
and decreasing $\log{g}$). Especially the macro-turbulent velocities for
changing $\teff$ agree well with the determined RHD velocities. 

\section{Summary}
We investigated the velocity fields in a set of M-star 3D hydrodynamical
models. They range in $\teff$ and $\log{g}$ from $\teff=2500$~K -- $4000$~K
and $\log{g}=3.0$ -- $5.0$ [cgs]. We applied a binning method to characterize
the velocity structure and to determine mean velocities. These velocities
range from $\sim 100$~m/s for the cool models up to $\sim 800 - 1000$~m/s for
hot models or models with small $\log{g}$ values.\\ In order to compare the 3D
RHD velocity fields with velocities needed in 1D spectral synthesis, we
expressed them in terms of classical micro- and macro-turbulent
velocities. For this purpose we computed the 3D and $<$3D$>$ \emph{Curve of
Growth} and determined the micro-turbulent velocity from them. They range in
the order of $\sim 100$~m/s for cool models to $\sim 600$~m/s for hot models
or models with small $\log{g}$. The macro-turbulent velocities were determined
through convolution with a Gaussian profile which, what turns out, is
sufficient in the regime of cool stars. The obtained velocities are of the
order of the 3D RHD velocity fields and show similar dependence from $\teff$
and $\log{g}$. We saw that 1D spectral synthesis in cool stars with micro-
and macro-turbulent velocity broadening is able to reproduce the 3D spectral
line synthesis.

\begin{theacknowledgments}
SW would like to acknowledge the support from the DFG Research Training Group
GrK - 1351 ``Extrasolar Planets and their host stars''.  AR acknowledges
research funding from the DFG under an Emmy Noether Fellowship (RE 1664/4- 1).
HGL acknowledges financial support from EU contract MEXT-CT-2004-014265
(CIFIST). We thank Derek Homeier for providing us with the opacity tables.

\end{theacknowledgments}

\bibliographystyle{aipproc}  

\bibliography{wende_boulder}

\end{document}